# High-throughput fabrication and semi-automated characterization of oxide thin film transistors

Yanbing Han[1,2], Sage Bauers[1], Qun Zhang[2*], Andriy Zakutayev[1*]

[1]Materials Science Center, National Renewable Energy Laboratory, Golden CO 80401, USA
[2]Department of Materials Science, Shanghai 200433, China

*Corresponding author: zhangqun@fudan.edu.cn (Qun Zhang);
andriy.zakutayev@nrel.gov (Andriy Zakutayev)

## Abstract

High throughput experimental methods are known to accelerate the rate of research, development, and deployment of electronic materials. For example, thin films with lateral gradients in composition, thickness, or other parameters have been used alongside spatially-resolved characterization to assess how various physical factors affect material properties under varying measurement conditions. Similarly, multi-layer electronic devices that contain such graded thin films as one or more of their layers can also be characterized spatially in order to optimize the performance. In this work, we apply these high throughput experimental methods to thin film transistors (TFTs), demonstrating combinatorial device fabrication and semi-automated characterization using sputtered Indium-Gallium-Zinc-Oxide (IGZO) TFTs as a case study. We show that both extrinsic and intrinsic types of device gradients can be generated in a TFT library, such as channel thickness and length, channel cation compositions, and oxygen atmosphere during deposition. We also present a semi-automated method to measure the 44 devices fabricated on a 50x50mm substrate that can help to identify properly functioning TFTs in the library and finish the measurement in a short time. Finally, we propose a fully automated characterization system for similar TFT libraries, which can be coupled with high throughput data analysis. These results demonstrate that high throughput methods can accelerate the investigation of TFTs and other electronic devices.

**TOC Figure:**

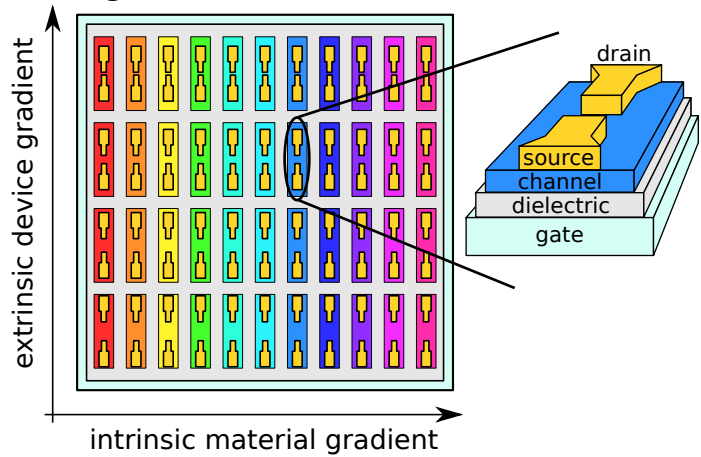

# 1. Introduction

Thin film transistors (TFTs) based on amorphous oxide semiconductors have recently received significant attention[1] due to their application in the control circuit for pixels in displays. When trying to optimize next-generation channel materials for their compositions[2], device sizes (like channel width/length[3], channel thickness[4]), or electrode metals[5], researchers have traditionally made several separate TFTs with different parameters, but this approach is both time and labor intensive. Due to the rapid development of scientific research and fierce competition in industry, a wide range of synthesis, characterization, and analysis parameters must be covered in a short time. High throughput methods[6,7] have been widely used in the investigation of material properties[8] and electronic devices.[9] It would be worthwhile to apply these high throughput methods towards the fabrication of TFTs to reduce the time needed optimize devices made from new materials.

There have been several examples where high throughput methods were applied to the investigation of oxide thin film transistors. For example, oxide TFT libraries with different In/Ga/Zn ratios were prepared to study the roles of each metallic element in Indium-Gallium-Zinc-Oxide (IGZO) TFTs[10]. As another example, $(ZnO)_x(SnO_2)_{1-x}$ TFTs were fabricated, with the highest electron mobilities observed at x≈0.25±0.05 and 0.80±0.03.[11] Effects of the In fraction on the devices performances and illumination instability of Indium Zinc Oxide (IZO) TFTs were investigated, showing that the mobility and subthreshold swing were improved as In increases.[12] Similar experiments have been attempted for Zn-In-Sn-O TFTs, with the best performance achieved at a composition ratio of Zn:In:Sn=40:20:40.[13] There are also reports on high throughput characterization, such as simultaneously testing of multiple TFT devices[14].

All prior combinatorial TFT studies focused on studying the performance as a function of composition metal elements in the oxide film. However, there are many other important parameters that influence the TFT performance in addition to metal compositions, such as device geometry (channel layer thickness and length), oxygen content in the film, placement of electrodes etc. For example, in the case of $CuSbS_2$ thin film photovoltaic solar cells, it has been shown that the crystallographic orientation and thickness of the active layer, as well as selection of metal for the back contact, have strong influence on the device performance.[15] It has been also found that the concentration of impurity elements diffusing from the glass can influence the device performance.[16] Thus more combinatorial work is needed to understand the influence of different synthesis, interfacial, and geometric parameters on TFT performance.

Here we report various types of device gradients that can be generated in a TFT library, such as channel thickness and length as well as chemical compositions of metals and oxygen. It is found that the channel thickness can be made uniform or non-uniform by depositing from two sources or one source at the same time. We also confirmed that both on-current and off-current decrease, as Gallium content and channel length increase. In addition, we showed that TFTs fabricated under oxygen rich conditions show lower off-current compared with the oxygen poor conditions. We utilized a semi-automated probe station to measure a TFT library that contains 44 TFT devices on a 50 x 50 mm substrate. This measurement helped determining the functional TFTs in one library and was considerably quicker than the single point measurement. Finally, we describe our proposal for a fully automated research-scale characterization tool for a TFT library and high-throughput analysis of the generated data. These results and proposals further advance high throughput fabrication, testing, and analysis method for TFTs.

## 2. Experimental details
### 2.1. Thin film deposition and characterization.

High throughput fabrication of the TFTs was based on the high throughput synthesis of thin films at NREL. Channel layer thin films were deposited onto 50 x 50 mm substrates (100 nm thermal $SiO_2$ on Si) using a combinatorial sputter chamber (AJA International, Orion 8). All the channel layers were deposited at room temperature (RT). Figure 1(a) is an illustration of the combinatorial sputtering chamber. All 4 sputter sources ("guns") have an incident angle of around 20 degrees to the substrate normal. Two identical Indium Zinc Oxide targets (IZO, In:Zn=63:37 wt. %) were mounted on Gun 1 and Gun 3. Sputtering power was kept at 40 W for all sputter guns, but it should be noted that this parameter can be adjusted to achieve a different composition or thickness range. Since the substrate is kept stationary during deposition, a thickness gradient would be presented in the sputtered thin films if only one IZO target was used. On the other hand, a uniform thickness could be obtained with two opposed IZO targets sputtered at the same time. A Gallium oxide ($Ga_2O_3$) target was mounted on Gun 2. By co-sputtering from Gun 1 (IZO) and Gun 2 ($Ga_2O_3$), a Ga composition gradient was expected with In:Zn ratio fixed as constant in one library. Similarly, another metal oxide target can be mounted on Gun 4 to act as a dopant or alloying agent in oxide TFT libraries.

The sputtering and reaction gases (Ar and $O_2$) also play an essential role in high throughput experiments. To this end, uniform and non-uniform gas atmospheres were introduced into the chamber. For uniform gas atmosphere, pure Ar flowed into the chamber through gas lines connected to the sputtering guns, as shown in Figure 1(b). The flow rate of pure Ar here was 5 sccm, which was split into 4 outlets at each of the 4 guns as shown in Figure 1(c), with the gas valve off. It is well known that in order to fabricate functional oxide thin film TFTs, oxygen gas should be introduced during sputtering to suppress oxygen vacancies[10]. A mixture of Ar and $O_2$ (Ar:$O_2$=95:5) was used as oxygen source. Ar/$O_2$ was introduced into the chamber with a flow rate of 10 sccm, and distributed into 6 even gas outlets by a gas distribution ring surrounding the substrate holder as shown in Figure 1(c). The sputtering pressure was 1.5 mTorr and the base pressure was below $10^{-6}$ Torr. The gas set up described above generated a uniform gas atmosphere relative to the substrate. On the other hand, non-uniform oxygen gas atmosphere was generated to investigate the oxygen influence on the TFT performance. Rather than being radially symmetric and introduced near the substrate surface, Ar/$O_2$ was instead connected to the gas line going to gun 3, while the rest of the set-up remained the same. This generated a local oxygen plasma-rich region near gun 3, thus leading to an oxygen content gradient in the thin films. Further information about similar combinatorial thin film deposition experiments can also be found in our previous reports.[17,18]

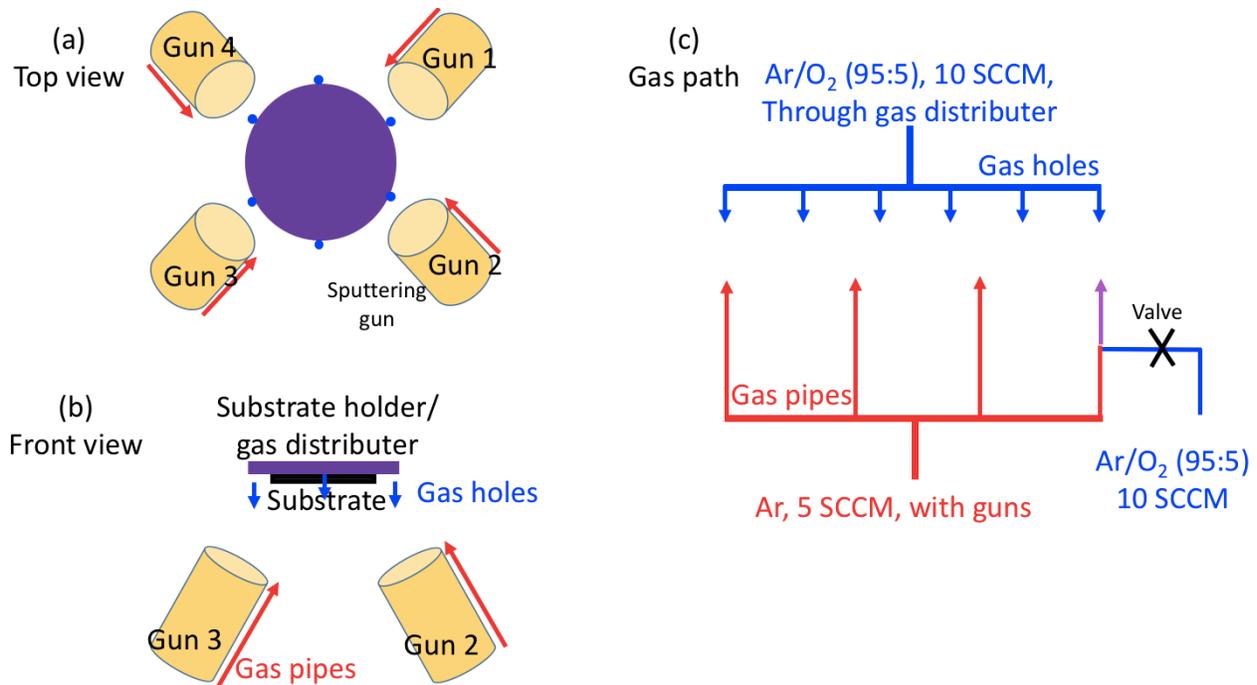

*Figure 1 Illustration of the combinatorial sputtering chamber used for channel layer growth. (a) Top view and (b) Front view of the chamber, showing sputter sources (guns) at an angle with respect to the substrate (c) Gas path of Ar and Ar/O$_2$, showing that the gas can be supplied uniformly near the substrate or non-uniformly near the sputter sources (guns)*

### 2.2 TFT device fabrication and testing.

Figure 2 shows the fabrication process for a combinatorial library of TFT devices. Devices were fabricated in a standard bottom-gate configuration, with a heavily doped Si wafer as the gate electrode, and 100 nm SiO$_2$ on its surface as a gate dielectric. In a region with no devices the SiO$_2$ was scratched away to expose the doped Si for contacting, which is a common method in the investigation of oxide TFTs.[19,20] The thin film channel layer was deposited as described above and patterned into 44 distinct areas by shadow mask, where each area was identified with a unique point number from 1 to 44. They are separated from each other to avoid any cross influences and to isolate any local regions with a leaky dielectric layer. The thickness and composition of the thin films were measured by X-ray fluorescence (XRF) on a Fischer XDV-SDD with a Ru X-ray source and a 3-mm diameter spot size. Aluminum source/drain electrodes with a uniform thickness of 300 nm were deposited on the channel layer via electron beam evaporation and also patterned using a shadow mask. The gradients in channel layer and electrode deposition on separate libraries resulted in gradients in the final devices. After the metallization, the TFT library was sent to furnace to be annealed in an air atmosphere at 300 °C. We found that normal TFT performance could only be obtained if the metallization was done before annealing; otherwise the gate voltage would not modulate channel resistance.

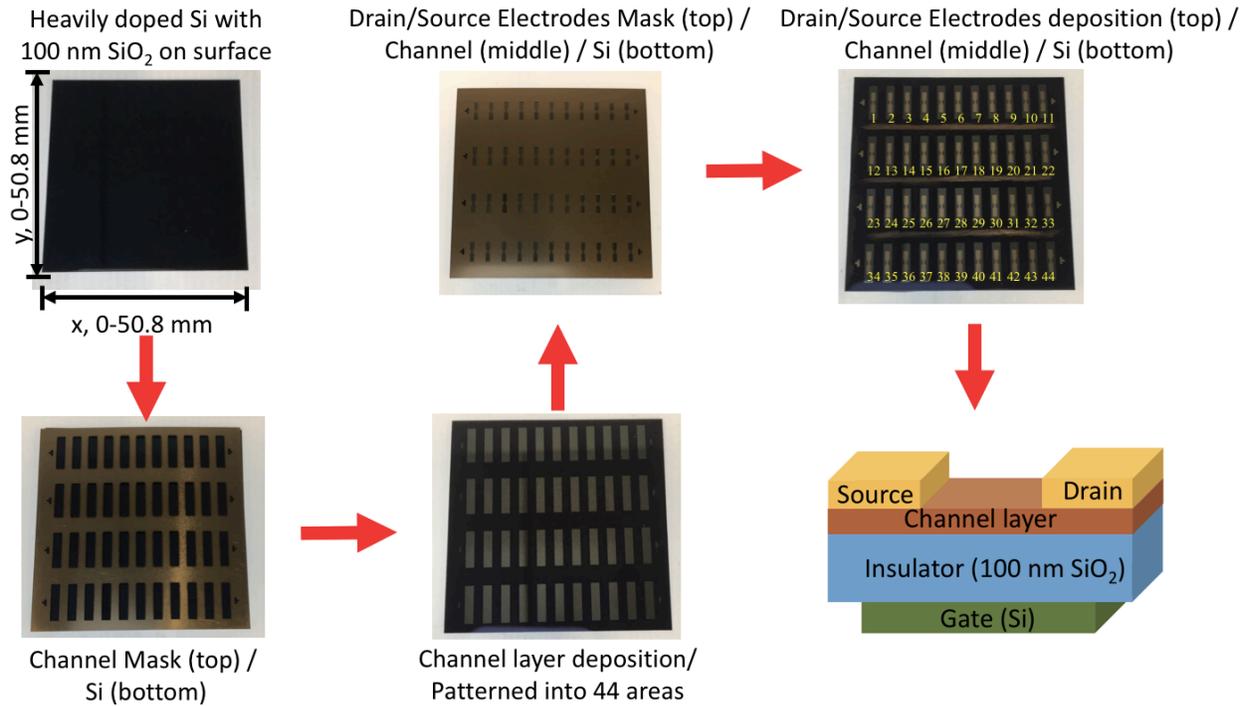

*Figure 2 Fabrication process of one TFT library containing 44 unique devices, staring with Si/SiO$_2$ stacks, and including two shadow mask steps for channel and source/drain contacts.*

High throughput characterization of a TFT device library is essential after high throughput fabrication. In these experiments, there are 44 devices in one library which would be cumbersome to measure if contacting the gate/source/drain electrodes one by one. While several tools are available for industrial-scale automated device testing over a wafer, they are now well suited to a research environment, because these instruments are large, expensive, and often only designed to accept a full-wafer of a particular size. Figure 3 shows the probe station used at NREL for semi-automated testing in this work. There are several adjustment knobs to control the sample stage, to move orthogonally along x-y coordinates, and to rotate. The control handle can set the "probe stage" that holds the three probe manipulators (for gate/source/drain electrodes) on two positions: the higher "hold position," and lower "measurement position". We have used this semi-automated method to reduce test time of a TFT library.

After aligning the sample so that rows/columns are along the movement directions, the measurement steps are as follows. (1) set the "probe stage" on "measurement position". (2) adjust the three probe manipulators to make sure the probe tips just touch the gate/source/drain electrodes from the 1$^{st}$ sample and conduct the characterization for this sample. (3) after the characterization, use the control handle to move up the "probe stage" to the "hold position". (4) since the probe manipulators are with the "probe stage", the probe tips will go up and leave the sample surface at the same time, but their relative positions in x-y plane won't change. (5) move the sample stage in x direction to the 2$^{nd}$ device. (6) use the control handle to move down the "probe stage" to the lower "measurement position". The probe tips will just touch the gate/source/drain electrodes since their relative positions stay the same. (7) conduct the test for 2$^{nd}$ device and repeat steps (1)-(6) to finish measurements for the other devices. Suggestions on how to automate the device measurement and data analysis processes further using COMBIgor[21] for Igor Pro are presented in SI.

This semi-automated measurement can finish 44 test in a relatively short time compared to using the micromanipulators. It also makes sure the probe tips touch the electrodes at the same position with the same probe pressure for each device, reducing the characterization error originating from the measurement conditions. Keithley 4200A-SCS Parameter Analyzer was utilized to measure the transfer curves for all the TFTs in this experiment. The gate voltage ($V_G$) was scanned from -10 V to +10 V with the drain voltage ($V_D$) fixed at 1 V to obtain transfer curve for all the devices. Since not all of the 44 devices show normal TFT performances due to the leakage in some areas of $SiO_2$, some measurements resulted in an abnormal high $I_G$. The transfer curves presented below are shown after excluding those with abnormal gate leakage current, and all the data are shown in Figure S1 in Supporting Information (SI).

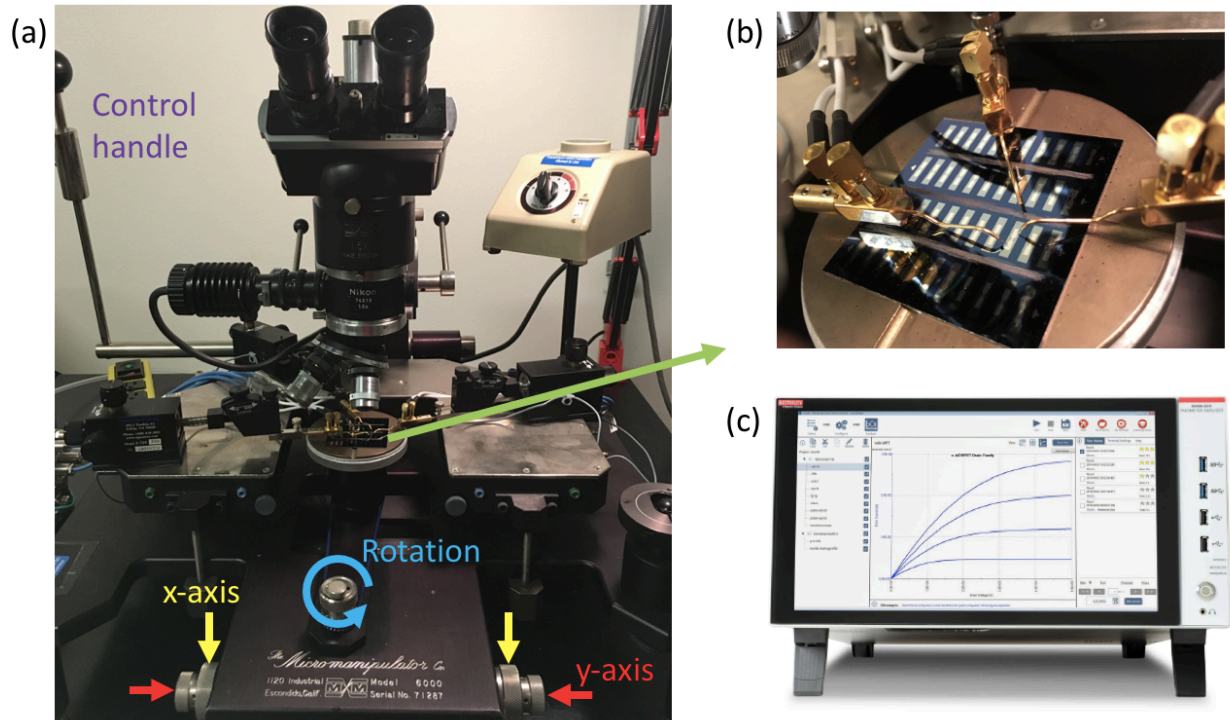

*Figure 3 Semi-automated measurement system for a library of TFTs, including (a) overview of the probe station, (b) focused view on the sample library under the probes, and (c) semiconductor parameter analyzers used for the measurement*

## 3. Results and discussion
### 3.1. Generation of the uniform and non-uniform channel thickness.

The channel thickness is a basic parameter in the investigation of TFTs, as well as for other electronic devices. As a baseline for the thickness-dependent and composition-dependent studies, we produced a uniform thickness in one library by placing two identical IZO guns at 180° angle with respect to each other (gun 1 and gun 3 in Figure 1). Figure 4(a) shows that the resulting thickness ranges from 102 to 108 nm after 60 minutes of sputtering and is expected to decrease to 51-54 nm with 30 minutes sputtering. The variation would be only 3 nm, which is likely low enough to avoid any obvious influence on the device performance when composition or other parameters are varied. An alternative route to uniform thickness films is substrate rotation during sputtering, which is not shown here. By fixing the thickness using rotation, one can study other parameters like the channel length.

To investigate the influences of the thickness on the device performance, a thickness gradient can be generated across the sample library when only one instead of two guns is used. The gradient would occur if the sputtering guns are tilted relative to the substrate normal, and the substrate is kept stationary during deposition. By only turning on the IZO target from gun 1, we fabricated IZO layers with an intentional thickness gradient, as shown in Figure 4(b). The IZO thickness ranges from 72 to 132 nm after 120 minutes sputtering. This result suggests that the thickness gradient can be applied in high throughput way to investigate its influence on the device performance. Using this result as a calibration point for the deposition rate of IZO TFT channel layers, we reduced the sputtering time to 30 minutes, which resulted in thicknesses in the 18-33 nm range. This thickness gradient (18-33 nm) didn't cause any obvious performance gradient in TFTs compared to the intrinsic variations among all the 44 devices in one library. Thus, developing methods to control the magnitude of the thickness gradient, such as changing the tilt angle of the sputter gun(s) or introducing moving shutters, would be required.

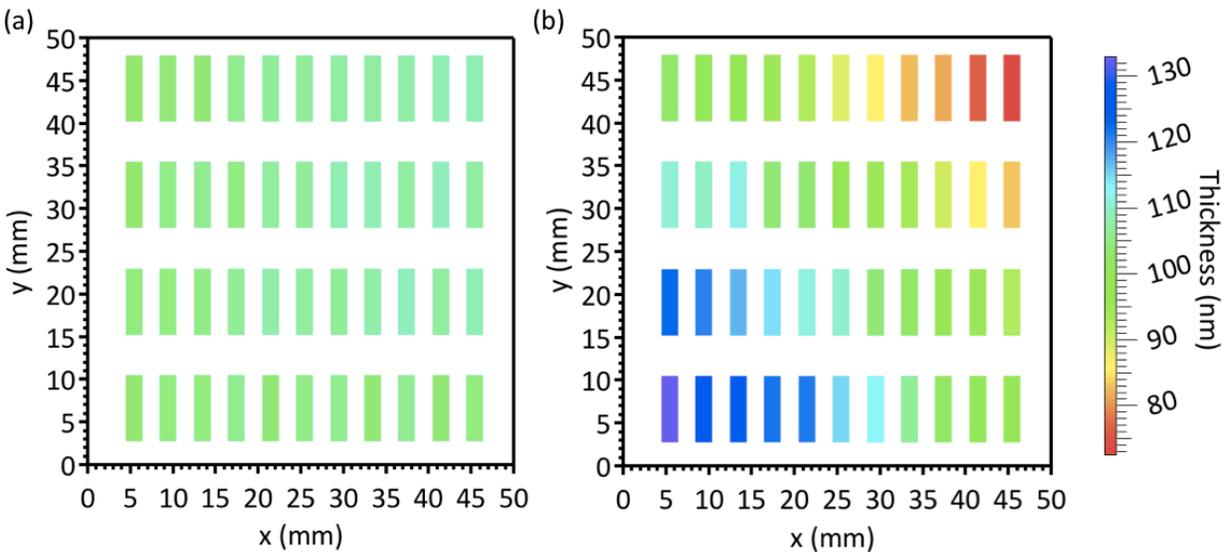

*Figure 4 (a) Uniform thickness can be obtained by sputtering from two same targets. (b) Thickness gradient generated in one sample library by using only one tilted gun.*

### 3.2. Composition gradients in high throughput fabrication of TFTs.

By placing two different targets in sputter guns opposite to each other, a composition gradient can be created across a sample library. Specifically, in this work, indium gallium zinc oxide (IGZO) channel layers were co-sputtered from a $Ga_2O_3$ target (Gun 1 in Figure 1) and an IZO target (Gun 2 in Figure 1). Deposition in this geometry results in a Ga/(Ga+In+Zn) atom ratio gradient as a function of position on the sample library as shown in Figure 5(a), while the relative In/Zn ratio is nominally fixed by the IZO target. Correspondingly, Figure 5(b) shows the transfer curves with different Ga compositions. Addition of Ga results in a higher threshold voltage ($V_{th}$) needed to turn on the channel, possibly due to suppression of oxygen vacancies in IGZO.[1]

Figure 5(c) summarizes the off-current and $V_{th}$ for transfer curves from Figure 5(b), where off-current is represented here by $I_D$ at Vth equals to -10 V and $V_{th}$ is defined as gate voltage at which $I_D$ reaches $10^{-9}$ A. Off-current decrease while $V_{th}$ increases as a function of Ga ratio, indicating lower carrier concentrations and higher resistance with higher Ga ratio. Previous combinatorial investigation of Ga/In/Zn oxide compositions utilized 3 targets, i.e. $Ga_2O_3$, $In_2O_3$, ZnO sputtering from 3 directions.[10] This helps to investigate metal compositions at the same time, but when Ga changes, In/Zn also shifts somewhat. In our work, by using IZO target, the Ga influence on IGZO TFTs with fixed In/Zn ratio can be observed.

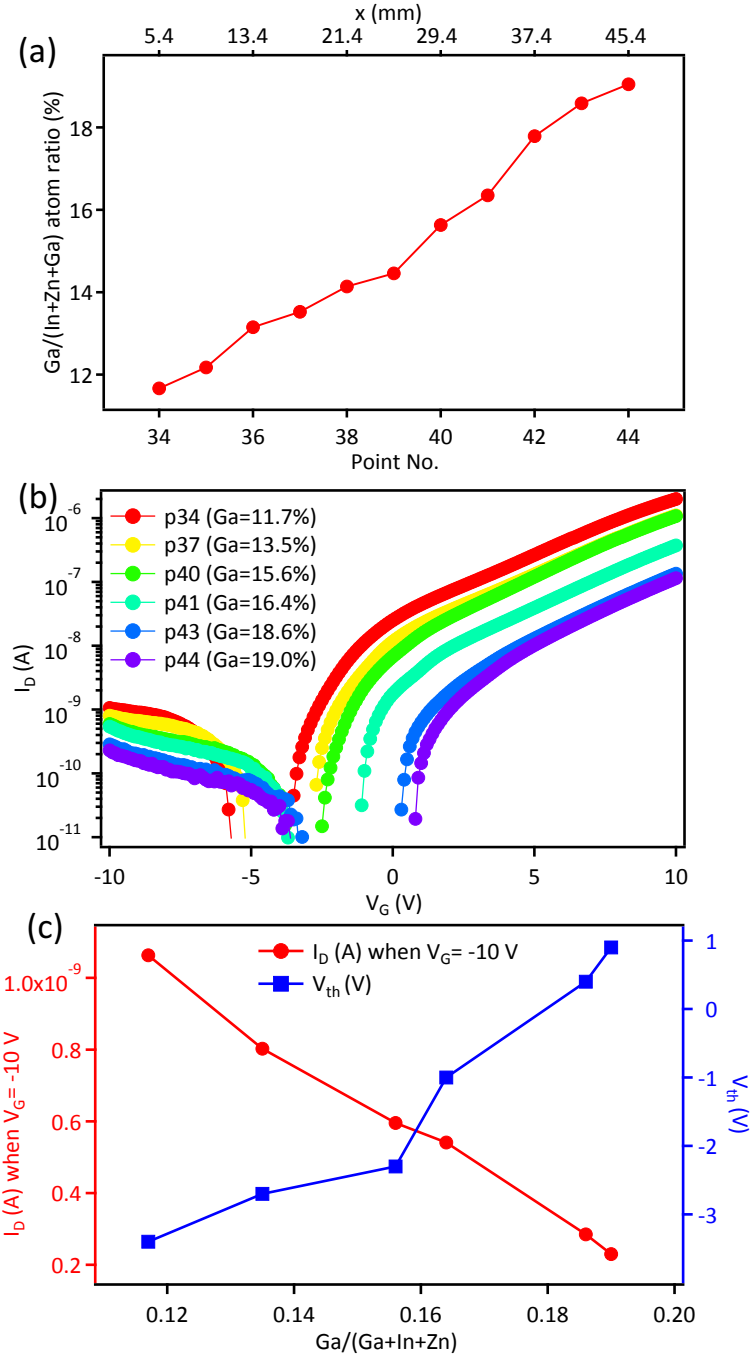

Figure 5 (a) Gallium gradient in high throughput TFTs, showing that Ga/(In+Zn+Ga) changes at different positions. (b) Transfer curves with different Ga ratios, showing change of performance with Ga content. (c) Off-current ($I_D$ when $V_G$=-10 V) decreases and $V_{th}$ increases when Ga ratio increases, suggesting decrease in carrier concentration and increase in the resistance of the channel.

### 3.3. Channel length gradient in high throughput fabrication of TFTs.

IZO TFTs with uniform thickness were prepared in order to compare channel length gradient. Source and Drain electrodes were patterned by metal deposition through stainless steel shadow masks with graded channel lengths. As shown in the inset of Figure 6(a), the channel length in row 2 and 3 varies from 0.1 mm to 3.6 mm, with a step of 0.35 mm. Channel length in row 4 is fixed at 0.1 mm, acting as a control group of devices. The center of the channel is kept at the same y-position in rows 2-4. On the other hand, the channel size in row 1 is the same for each x-position as in rows 2-3, but one row of electrodes is at a fixed y-position. Their positions are not symmetric on the IZO film to investigate possible influences of asymmetric electrode placement on the underlying channels.

Figure 6(a) displays the transfer curves of several devices from rows 2-3 of the device library shown in the inset of Figure 6(a). Transfer curves from Point 37, 38, 39, 41 in Row 4 are also plotted to assess variability across nominally identical devices. The 4 transfer curves from Row 4 overlap with each other in the on-state region ($V_G > 0$ V), but show some variance in off-state region ($V_G < 0$ V). This suggests that any differences of on-current in devices from rows 2 and 3 will be due to changes in channel length as opposed to inherent device variability. Generally, as the channel length increases, the devices show a larger $V_{th}$. This can be attributed to the higher resistance with longer channel, since it takes a higher voltage to gather more carriers to reach the same current level to turn on the TFT.

Figure 6(b) shows $I_D$ when $V_G = +10$ V (on-state of the transistor) as a function of channel length based on the transfer curves in Figure 6(a). This quantity is proportional to on/off ratio of the TFT, if $I_D$ in the off-state does not vary. However, in our case differences for $I_D$ when $V_G = -10$ V (off-state of the transistor) are non-negligible due to intrinsic variation in the off-state region. On the other hand, the variation for $I_D$ when $V_G = +10$ V is negligible for TFTs with channel length 0.1 mm, as shown in Figure 6(a). As the channel length increases, $I_D$ decreases since resistance becomes larger, leading to a lower on-current in Figure 6(b). It is interesting to note that the slope of this dependence changes below channel length of 1 mm, which may be related to "pinch-off" effects in the transistor. Finally, we didn't notice any obvious influences induced by non-symmetric electrodes after comparing transfer curves between Row 1 and Row 2 or Row 3. The results here show that a designed channel length gradient can lead to obvious gradient in transfer curves, which allows us to investigate the channel length influences in a high-throughput way.

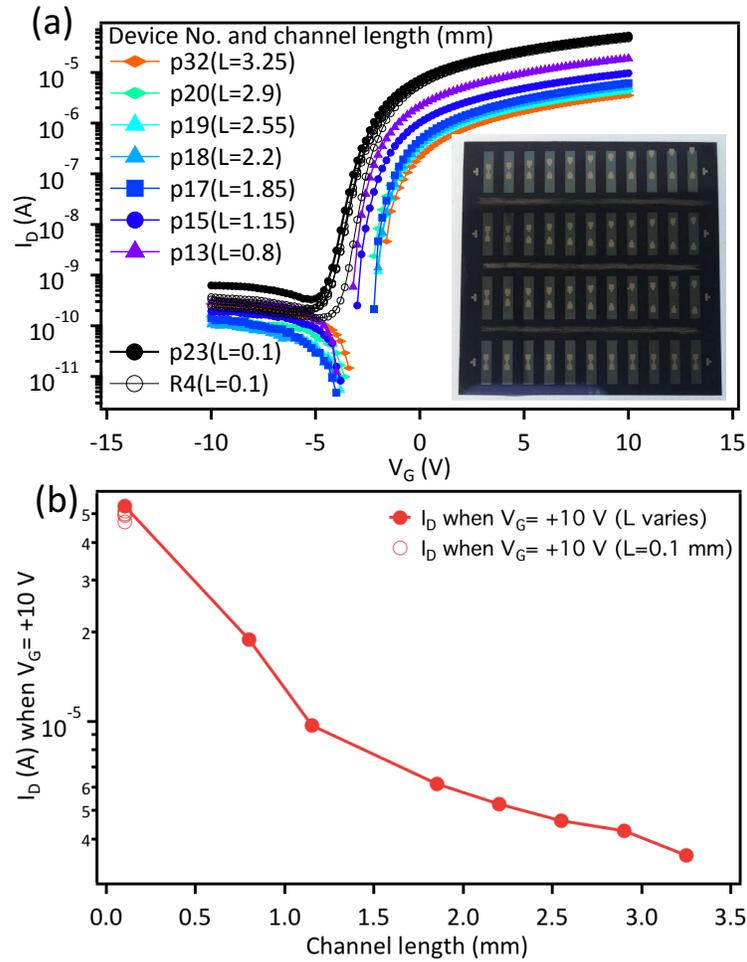

Figure 6 (a) Transfer curves from TFT library with different channel lengths. The inset is TFT library with different channel lengths. (b) $I_D$ when $V_G = +10$ V as a function of channel length. Longer channels lead to a lower on-current.

### 3.4. Non-uniform oxygen atmosphere phenomenon in high throughput fabrication of TFTs.

Previously we reported on sputtering with combinatorial gas gradients, showing that a nitrogen doping gradient can be introduced into sputtered thin films by carefully selecting the gas outlet positions in the vacuum chamber.[18] Here we used a similar approach to create oxygen gradients in IZO thin films as explained in Experimental Details above. Gun 1 is IZO target with pure Ar while Gun 3 is IZO with $Ar/O_2$ as sputtering gas. For this analysis, we simply divided the 44 points into two parts as shown in Figure 7(a). The top right half is an oxygen rich region since it is near Gun 3 and the other half is oxygen poor region since it is near Gun 1.

Figure 4(b) shows transfer curves from the oxygen poor region in red and oxygen rich region in green. It is interesting to note that generally transfer curves from the oxygen poor area show a higher current than those from the oxygen rich area in the off-state, while they are similar in the on-state. One hypothesis to explain this phenomenon is that oxygen poor regions have higher density of oxygen vacancies, which have been claimed to be the origin of mobile carriers in IZO. However, further measurement and quantification of oxygen content, oxygen valence states, and carrier concentrations would be required to confirm this hypothesis.

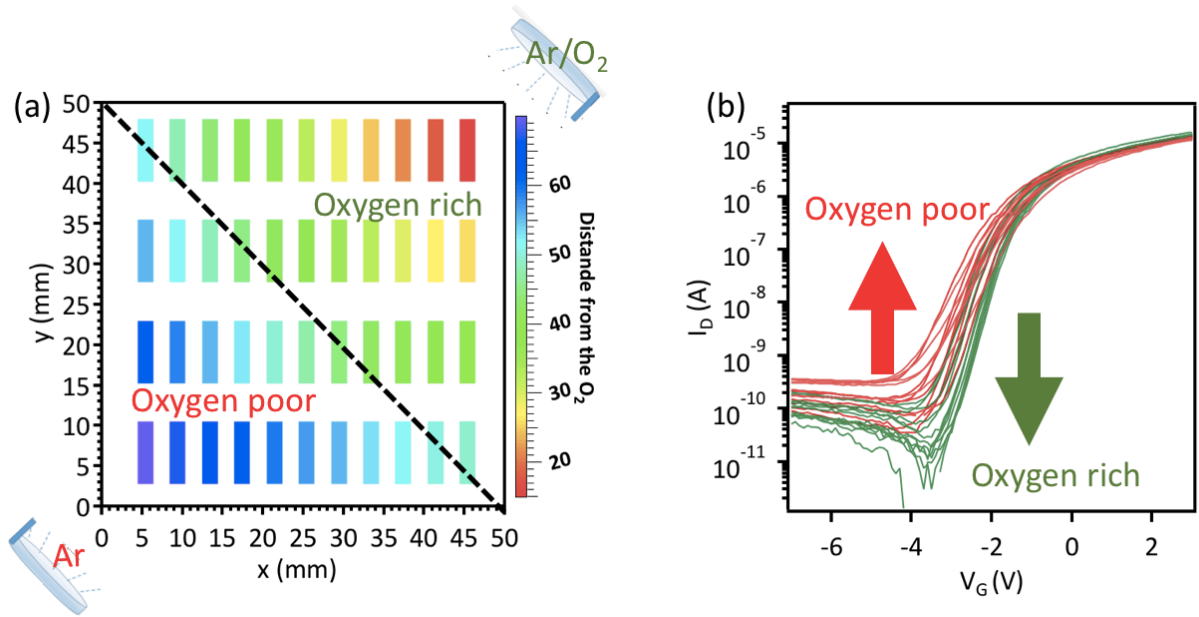

*Figure 4 (a) Sputtering guns (IZO with Ar, IZO with Ar/O$_2$) positions relative to the substrate and sample positions relative to the oxygen source. (b)Transfer curves in high throughput TFT fabricated from non-uniform oxygen atmosphere.*

## 4. Summary

In this paper, we reported on a case-study of high-throughput fabrication and analysis of TFTs at NREL, and discussed our future plans of fully automated testing and rapid analysis of TFT libraries in supporting information. We showed that high-throughput methods can be applied to determine how both intrinsic (material) and extrinsic (device) parameters, such as the composition, channel thickness, channel size, and oxygen content affect performance of oxide TFTs. For the IGZO TFTs studied here, we found that the off-current decreases and threshold voltage increases with increasing Ga content, confirming that Ga can suppress the electron concentration in IZO. In addition, we showed that on-current decreases and threshold voltage becomes more positive as channel length increases, since the longer channel leads to higher resistance. It is also shown that addition of oxygen can decrease the channel current, as observed in the oxygen rich region of one TFT library. All these high-throughput experimental results help to cover a wide range of fabrication conditions for TFTs in relatively short time, which is needed to improve device performance. The results presented here demonstrate how high-throughput methods can accelerate the investigation of TFTs and other electronic devices.

## Author information
Yanbing Han, Email address: ybhan14@fudan.edu.cn
Sage Bauers, Email address: Sage.Bauers@nrel.gov
Qun Zhang, Email address: zhangqun@fudan.edu.cn
Andriy Zakutayev, Email address: Andriy.Zakutayev@nrel.gov

## Acknowledgement

This work was supported by the U.S. Department of Energy under Contract No. DE-AC36-08GO28308 with Alliance for Sustainable Energy, LLC, the Manager and Operator of the National Renewable Energy Laboratory (NREL). Funding provided by Laboratory Directed Research and Development (LDRD) program at NREL. Y. H. acknowledges the support from Science and Technology Commission of Shanghai Municipality (No. 16JC1400603), and a grant from the National Natural Science Foundation of China (No. 61471126). Y. H. also thanks the China Scholarship Council for offering the stipend to perform research at the National Renewable Energy Laboratory. The views expressed in the article do not necessarily represent the views of the DOE or the U.S. Government.